\documentclass[11pt]{amsart}

\usepackage[utf8]{inputenc}
\usepackage{amssymb}
\usepackage{epsfig}
\usepackage{comment}
\usepackage{amsmath}
\usepackage[section]{placeins}
\usepackage{mathrsfs}
\usepackage{graphicx}
\usepackage[notrig]{physics}
\usepackage{subfigure}
\usepackage{caption}
\usepackage{color}
\usepackage{svg}

\graphicspath
{
  {./}
  {./Figures/}
}

\begin{document}

\title{Fracture and Size Effect in Mechanical Metamaterials}
\author{
J.~Ulloa${}^1$, M.~P.~Ariza${}^2$, J.~E.~Andrade${}^1$ and M.~Ortiz${}^{1}$
}

\address
{
 ${}^1$Division of Engineering and Applied Science, California Institute of Technology, 1200 E.~California Blvd., Pasadena, CA 91125, USA.
  \\
 ${}^2$Escuela T\'ecnica Superior de Ingenier\'ia, Universidad de Sevilla, Camino de los descubrimientos, s.n., 41092 Sevilla, Spain.
 }

\email{$\dots$}

\begin{abstract}
We resort to variational methods to evaluate the asymptotic behavior of fine metamaterials as a function of cell size. To zeroth order, the metamaterial behaves as a micropolar continuum with both displacement and rotation degrees of freedom, but exhibits linear-elastic fracture mechanics scaling and therefore no size effect. To higher order, the overall energetics of the metastructure can be characterized explicitly in terms of the solution of the zeroth-order continuum problem by the method of $\Gamma$-expansion. We present explicit expressions of the second-order correction for octet frames. As an application, we evaluate the compliance of double-cantilever octet specimens to second order and use the result to elucidate the dependence of the apparent toughness of the specimen on cell size. The analysis predicts the discreteness of the metamaterial lattice to effectively shield the crack-tip, a mechanism that we term {\sl lattice shielding}. The theory specifically predicts {\sl anti-shielding}, i.~e., {\sl coarser is weaker}, in agreement with recent experimental observations.
\end{abstract}

\maketitle

\section{Introduction}

The mechanical properties of metamaterials have been extensively analyzed by a variety of means, experimental, computational  and analytical (cf. \cite{Ashby2006, Fleck2010,Montemayor2015,kochmann:2019,Greer:2019,BENEDETTI2021,LU2022,Jiao2023,Jin2024} for reviews and background). A case of particular interest in practice concerns fine metastructures in which the cell size is much smaller than the size of the domain. Under such conditions, it is reasonable to expect that the elastic field of the metastructures can be approximated by some continuum elastic field, to be determined, the approximation becoming increasingly accurate for increasingly smaller unit-cell sizes. Thus understood, the continuum limit of metastructures is a particular case of {\sl discrete-to-continuum} limits in the calculus of variations. Discrete-to-continuum methods of analysis were pioneered by \cite{Alicandro:2004} and \cite{Braides:2004} and have since been applied to a wide range of problems.

The overwhelming advantage of discrete-to-continuum analysis is that the form itself of the limiting field, and not just material parameters thereof, is determined unambiguously by the analysis. In the case of metamaterials, the analysis shows that, to zeroth order in the cell size, the continuum limit of metamaterial frames, undergoing both axial and bending deformations, is {\sl micropolar} \cite{Ariza:2024}, in the sense of \cite{Eringen:1968}, Section~\ref{uPa8Ag}. The analysis also provides closed-form expressions for the elastic and polar moduli as a function of the geometry of the metamaterial and the axial and bending stiffnesses of the bars \cite{Ariza:2024}. In separate work, we have assessed computationally the convergence of discrete solutions to the continuum limit over a range of two and three-dimensional test cases \cite{Ulloa:2024}. The calculations verify the analytical results and exhibit strong---albeit configuration-dependent---rates of convergence.

These convergence properties notwithstanding, the zeroth-order discrete-to-continuum limit turns out to be uninformative in a number of important respects, such as the {\sl size effect} observed experimentally in fracture tests of metamaterials \cite{Shaikeea2022}. Indeed, the zeroth-order discrete-to-continuum analysis entirely wipes out any notion of cell size from the limit, and the resulting behavior---albeit micropolar---exhibits the scaling of linear elasticity and fails to capture size effects. Situations where variational limits are not sufficiently informative arise with some frequency in multiscale analysis, which has motivated a number of extensions designed to retain unit-cell level information to various orders when passing to the continuum limit \cite{Braides:2008}.

In this work, we resort to the method of $\Gamma$-expansion of \cite{Anzellotti:1993} in order to extend the discrete-to-continuum analysis of metamaterials to higher order. As applied here, the method of $\Gamma$-expansion generates an expansion for the overall energetics of the system to arbitrary order in the cell size whose evaluation conveniently requires only the solution of the problem to zeroth order. Using this approach, we derive explicit second-order corrections for octet metamaterials.

As an example of application, we seek to elucidate the size-dependence of the apparent toughness of metamaterials in fracture tests \cite{Shaikeea2022} using compliance analysis \cite{Hutchinson:1979,Luan2022}. To this end, we consider an ideal double-cantilever specimen (cf.~\cite{Maurizi2022} for the use of compact-tension specimens for testing metamaterials) and evaluate its elastic compliance to second order in the cell size. We verify the excellent accuracy of the predicted compliances by comparison to direct discrete calculations. Compliance analysis \cite{Hutchinson:1979} then supplies the requisite energy-release rate, which is explicitly dependent on cell size in the second-order expansion. When combined with a Griffith brittle fracture criterion for the failure of the bars, the analysis predicts an apparent toughness that depends on the ratio of the cell size to the size of the specimen. This effect is in analogy to crack-tip shielding in composites, where the effective toughness of the composite is controlled by the net energy-flux to the crack tip, as measured by the $J$-integral (cf., e.~g., \cite{Hutchinson:1987}). We therefore refer to the effect of the discreteness of the lattice on the apparent toughness of the metamaterial as {\sl lattice shielding}. More specifically, the analysis predicts that increasing the unit-cell size, or coarsening the metamaterial, results in a reduced apparent toughness of the specimen, i.~e., it results in {\sl anti-shielding}. These trends agree qualitatively with the experimental observations and measurements of \cite{Shaikeea2022}, which provides a modicum of validation of the theory.

\section{Homogenization and continuum limit}

We consider a sequence of increasingly finer metamaterials, scaled by a small parameter $\epsilon > 0$, contained within a fixed domain $\Omega$ and deforming under the action of fixed loads $f(x) = \big(q(x) ;\, m(x) \big)$ per unit volume independent of $\epsilon$, where $q(x)$ and $m(x)$ are distributed forces and moments, or torques, per unit volume, respectively. We endeavor to ascertain the asymptotic behavior of the metamaterials in the limit of $\epsilon \to 0$, or {\sl continuum limit}.

It should be carefully noted that the analysis is limited to linear elasticity, with quadratic energy forms as the point of departure. Non-linear effects such as buckling of the bars under compression, which result in loss of rank-one convexity~(\cite{muller1993}), are out of scope. However, the analysis does take into account linearized joint rotations and beam bending in the framework of elementary beam theory.

\subsection{Scaling}

In order to define a corresponding sequence of scaled energies, we designate a reference metamaterial of fixed lattice size and denote by $E(u; \Omega/\epsilon)$ the energy of the corresponding metastructure maximally contained in $\Omega/\epsilon$. The energies of the scaled metastructures maximally contained in $\Omega$ are, then,
\begin{equation} \label{R79mPV}
    E_\epsilon(u_\epsilon; \Omega)
    =
    \epsilon^{n-2} \, E(u; \Omega/\epsilon) ,
\end{equation}
where $u = (v;\theta)$ collects the displacements and rotations defined over the reference metamaterial spanning the domain $\Omega/\epsilon$ and $u_\epsilon = (\epsilon v;\theta)$ are scaled displacements and rotations over the scaled metastructure spanning the domain $\Omega$. The corresponding sequence of potential energies is, then,
\begin{equation} \label{1C3ijx}
    F_\epsilon(u_\epsilon; \Omega)
    =
    E_\epsilon(u_\epsilon; \Omega)
    -
    \langle f,\, u_\epsilon \rangle ,
\end{equation}
where the last term represents the negative work of the applied forces.

We note that the preceding scaling is chosen so that both strains and rotations remain $O(1)$ as $\epsilon \downarrow 0$. In addition, the factor $\epsilon^{n-2}$ accounts for the expected elasticity scaling of the limiting energy and is included to ensure a proper limit \cite{Espanol:2013, Ariza:2024}.

\subsection{Heuristics} \label{QIcmVs}

We may regard the potential energy (\ref{1C3ijx}) of the scaled metastructure symbolically as a quadratic form,
\begin{equation}
    F_\epsilon(u)
    =
    \frac{1}{2}
    \mathbb{K}_\epsilon u \cdot u
    -
    f^T u ,
\end{equation}
where $\mathbb{K}_\epsilon$ represents the stiffness matrix of the metastructure at the $\epsilon$ scale and we omit reference to $\Omega$ for simplicity of notation. We denote by $u_\epsilon$ the minimizer of $F_\epsilon$,
\begin{equation}
    F_\epsilon(u_\epsilon)
    =
    \min F_\epsilon
    =
    m_\epsilon
    =
    -
    \frac{1}{2}
    \mathbb{K}_\epsilon^{-1}
    f \cdot f ,
\end{equation}
representing the stable equilibrium displacement of the metastructure.

We seek an asymptotic expansion of the form
\begin{equation} \label{9UWwP1}
    F_\epsilon \sim F_0 + \epsilon F_1 + \dots + \epsilon^s F_s + o(\epsilon^s) ,
\end{equation}
for some energy functions $F_0,\dots,F_s$, to be defined. We specifically follow the variational theory of $\Gamma$-expansions of \cite{Anzellotti:1993}, which ensures proper converge of the $\epsilon$ minimizers, as well as convergence of the energy and any continuous quantity of interest. In this approach, we seek an expansion of the minimum energy of the form
\begin{equation} \label{bzynWH}
    m_\epsilon \sim m_0 + \epsilon m_1 + \dots + \epsilon^s m_s + o(\epsilon^s) .
\end{equation}

To zeroth order,
\begin{equation}
    m_0
    =
    \lim_{\epsilon\to 0} m_\epsilon
    =
    -
    \frac{1}{2}
    \mathbb{K}_0^{-1}
    f \cdot f ,
\end{equation}
where
\begin{equation}
    \mathbb{K}_0
    =
    \Big(
        \lim_{\epsilon\to 0} \mathbb{K}_\epsilon^{-1}
    \Big)^{-1}
\end{equation}
is the zeroth-order continuum operator. The corresponding potential energy is the quadratic form
\begin{equation}
    F_0(u)
    =
    \frac{1}{2}
    \mathbb{K}_0 u \cdot u
    -
    f^T u ,
\end{equation}
which defines the zeroth-order term in the expansion (\ref{9UWwP1}).
The minimizer $u_0$, with the property
\begin{equation}
    F_0(u_0)
    =
    \min F_0
    =
    m_0 ,
\end{equation}
characterizes the stable equilibrium of the metastructure in the continuum limit. We note that
\begin{equation}
    u_0 = \lim_{\epsilon\to 0} u_\epsilon ,
    \quad
    m_0 = \lim_{\epsilon\to 0} m_\epsilon ,
\end{equation}
i.~e., $\Gamma$-expansion ensures the convergence of minimizers and energies.

To first order,
\begin{equation}
    m_1
    =
    \lim_{\epsilon\to 0}
    \frac{m_\epsilon - m_0}{\epsilon}
    :=
    -
    \frac{1}{2}
    \mathbb{K}_1^{-1}
    f \cdot f ,
\end{equation}
where
\begin{equation}
    \mathbb{K}_1
    =
    \Big(
        \frac{d}{d\epsilon}
        \Big(
            \mathbb{K}_\epsilon^{-1}
        \Big)_{\epsilon=0}
    \Big)^{-1}
    =
    -
    \Big(
        \mathbb{K}_0^{-1}
        \mathbb{K}'_0
        \mathbb{K}_0^{-1}
    \Big)^{-1}
\end{equation}
is the first-order continuum operator, where we denote $$\mathbb{K}'_0=\bigg(\frac{d}{d\epsilon}\mathbb{K}_\epsilon\bigg)_{\epsilon=0}.$$ The corresponding potential energy is the quadratic form
\begin{equation}
    F_1(u)
    =
    \frac{1}{2}
    \mathbb{K}_1 u \cdot u
    -
    f^T u ,
\end{equation}
which defines the first-order term in the expansion (\ref{9UWwP1}). We note that
\begin{equation}
    m_1
    =
    \frac{1}{2}
    \Big(
        \mathbb{K}_0^{-1}
        \mathbb{K}'_0
        \mathbb{K}_0^{-1}
    \Big)
    f \cdot f
    =
    \frac{1}{2}
    \mathbb{K}'_0 u_0 \cdot u_0 ,
\end{equation}
i.~e., the first order correction $m_1$ of the energy at equilibrium is determined directly by the zeroth-order continuum displacements $u_0$.

To second order, we have
\begin{equation}
    m_2
    =
    \lim_{\epsilon\to 0}
    \frac{m_\epsilon - m_0 - \epsilon m_1}{\epsilon^2}
    :=
    -
    \frac{1}{2}
    \mathbb{K}_2^{-1}
    f \cdot f ,
\end{equation}
where
\begin{equation}
    \mathbb{K}_2
    =
    \Big(
        \frac{1}{2}
        \frac{d^2}{d\epsilon^2}
        \Big(
            \mathbb{K}_\epsilon^{-1}
        \Big)_{\epsilon=0}
    \Big)^{-1} .
\end{equation}
If, in addition $\mathbb{K}'_0=0$, then
\begin{equation}
    \mathbb{K}_2
    =
    -
    \Big(
        \frac{1}{2}
        \mathbb{K}_0^{-1}
        \mathbb{K}''_0
        \mathbb{K}_0^{-1}
    \Big)^{-1} ,
\end{equation}
whereupon
\begin{equation}
    m_2
    =
    \Big(
        \frac{1}{4}
        \mathbb{K}_0^{-1}
        \mathbb{K}''_0
        \mathbb{K}_0^{-1}
    \Big)
    f \cdot f
    =
    \frac{1}{4}
    \mathbb{K}''_0 u_0 \cdot u_0 ,
\end{equation}
which is again determined directly by the zeroth-order continuum displacements $u_0$.

Higher-order terms in the $\Gamma$-expansion follow by recursion. A rigorous version of the preceding plan was put forth in the seminal paper of \cite{Anzellotti:1993} under the name of $\Gamma$-expansion, and has become a standard staple in the calculus of variations. In $\Gamma$-expansion, the energy functions $F_0,\dots,F_s$ in (\ref{9UWwP1}) are defined recursively by $\Gamma$-convergence (cf., e.~g., \cite{dalmaso:1993} for background), which in general entails relaxation in the form of weakly convergent local oscillations. By virtue of this relaxation, the energy functionals $F_0,\dots,F_s$ may undershoot significantly naive pointwise limits. In the case of metastructures, relaxation takes place, for instance, in the form of energy-minimizing displacements of the joints at the unit cell level, especially when the metastructure contains joints of different types \cite{Ariza:2024}.

\subsection{Zeroth-order homogenization} \label{uPa8Ag}

The zeroth-order limit $E_0$ in the $\Gamma$-expansion (\ref{9UWwP1}) coincides with the $\Gamma$-limit of the sequence $(E_\epsilon)$ of scaled energies as $\epsilon \to 0$. This limit is greatly simplified in the case of linear metastructures, wherein the total energy is a quadratic form (see, e.~g., \cite{dalmaso:1993} for background on $\Gamma$ and $G$-convergence of quadratic forms). The result is \cite{Ariza:2024}
\begin{equation} \label{lIKdde}
    E_0(u_0)
    =
    \int_{\Omega}
        W_0\big(\varepsilon_0(x),\, \theta_0(x) - \frac{1}{2} \operatorname{curl} v_0(x)\big)
    \, dx ,
\end{equation}
where $\varepsilon_0(x) = {\rm sym} Dv_0(x)$ are the local strains and the quadratic function $W_0(\varepsilon,\, \omega)$, $(\varepsilon,\, \omega) \in \mathbb{R}^{n\times n} \times \mathbb{R}^{n(n-1)/2}$, is the effective energy density of the infinite metamaterial.

Thus, to lowest order, the effective continuum energy (\ref{lIKdde}) of linear metastructures is a special case of linear {\sl micropolar elasticity}, in the sense of \cite{Eringen:1966}, in which the energy density is independent of the curvature, or bending strain, $D\theta_0(x)$. Explicit expressions for the attendant effective moduli of two-dimensional honeycomb lattices and three-dimensional octet trusses are presented in \cite{Ariza:2024}.

\subsection{Higher-order homogenization}

The zeroth-order effective continuum energy (\ref{lIKdde}) is {\sl local} in the sense of volume scaling, i.~e., $E_0(u'_0; \lambda^{-1}\Omega) = \lambda^2 E_0(u_0; \Omega)$ if $u'_0(x) = u_0(\lambda^{-1} x)$. Therefore, the zeroth-order limit fails to capture size effects such as are observed at the mesoscale \cite{Shaikeea2022}. In order to retain mesoscopic information in the theory, we extend the asymptotic analysis to terms of higher order in $\epsilon$. To this end, we follow the $\Gamma$-expansion program sketched out in Section~\ref{QIcmVs}.

Suppose that the metamaterial is loaded by macroscopic forces $f(x)$. Then, in Fourier representation, the equilibrium equations take the form \cite{Ariza:2024}
\begin{equation} \label{ZqVspb}
    \mathbb{D}_\epsilon(k) \, \hat{u}_\epsilon(k)
    =
    \hat{f}(k) ,
    \quad
    k \in B/\epsilon ,
\end{equation}
where $\mathbb{D}_\epsilon(k)$ and $B/\epsilon$ are the dynamical matrix and Brillouin zone of the scaled metamaterial, $\hat{u}_\epsilon(k)$ and $\hat{f}(k)$ are the discrete Fourier transform of displacements and forces. Provided that the matrix $\mathbb{D}_\epsilon(k)$ is non-singular, which is a requirement of structural stability, the equilibrium problem (\ref{ZqVspb}) can be solved pointwise, with the result
\begin{equation} \label{jQ2xKs}
    \hat{u}_\epsilon(k)
    =
    \mathbb{D}_\epsilon^{-1}(k) \, \hat{f}(k) ,
    \quad
    k \in B/\epsilon .
\end{equation}
The corresponding minimum potential energy is
\begin{equation}\label{meps}
    m_\epsilon
    =
    -
    \frac{1}{(2\pi)^n}
    \int_{B/\epsilon}
        \frac{1}{2}
        \mathbb{D}_\epsilon^{-1}(k)
        \, \hat{f}(k) \cdot \hat{f}^*(k)
    \, dk .
\end{equation}
Passing to the limit $\epsilon\to 0$ gives the minimum energy to zeroth order, namely \cite{Ariza:2024},
\begin{equation}
    m_0
    =
    -
    \frac{1}{(2\pi)^n}
    \int_{\mathbb{R}^n}
        \frac{1}{2}
        \mathbb{D}_0^{-1}(k)
        \, \hat{f}(k) \cdot \hat{f}^*(k)
    \, dk .
\end{equation}

In order to evaluate higher-order corrections in energy, we suppose that the macroscopic loads $f(x)$ are {\sl slowly-varying}, in the sense that $\hat{f}(k)$ is compactly supported.

To first order in the expansion (\ref{bzynWH}), we have
\begin{equation}
    m_1
    =
    \lim_{\epsilon\to 0}
    \frac{m_\epsilon - m_0}{\epsilon} ,
\end{equation}
or
\begin{equation}
    m_1
    =
    -
    \lim_{\epsilon\to 0}
    \frac{1}{(2\pi)^n}
    \int_{B/\epsilon}
        \frac{1}{2}
        \frac{1}{\epsilon}
        \Big(
            L^T
            \mathbb{D}_\epsilon^{-1}(k)
            L
            -
            \mathbb{D}_0^{-1}(k)
        \Big)
        \, \hat{f}(k) \cdot \hat{f}^*(k)
    \, dk .
\end{equation}
Passing to the limit,
\begin{equation}
    m_1
    =
    -
    \frac{1}{(2\pi)^n}
    \int_{B/\epsilon}
        \frac{1}{2}
        \frac{d}{d\epsilon}
        \Big(
            L^T
            \mathbb{D}_\epsilon^{-1}(k)
            L
        \Big)_{\epsilon=0}
        \, \hat{f}(k) \cdot \hat{f}^*(k)
    \, dk ,
\end{equation}
which is the first-order correction to the energy of the metamaterial at equilibrium. We note that $m_1$ vanishes if the metamaterial is centrosymmetric, i.~e., if $\mathbb{D}_\epsilon(-k) = - \mathbb{D}_\epsilon(k)$, in which case size effects are second order in $\epsilon$ or weaker.

To second order, we likewise have
\begin{equation}
    m_2
    =
    \lim_{\epsilon\to 0}
    \frac{m_\epsilon - m_0 - \epsilon m_1}{\epsilon^2} ,
\end{equation}
or, assuming $\mathbb{D}'_0(k) = 0$ for simplicity,
\begin{equation}
    m_2
    =
    -
    \lim_{\epsilon\to 0}
    \frac{1}{(2\pi)^n}
    \int_{B/\epsilon}
        \frac{1}{2}
        \frac{1}{\epsilon^2}
        \Big(
            L^T
            \mathbb{D}_\epsilon^{-1}(k)
            L
            -
            \mathbb{D}_0^{-1}(k)
        \Big)
        \, \hat{f}(k) \cdot \hat{f}^*(k)
    \, dk .
\end{equation}
Passing to the limit,
\begin{equation}\label{m2N}
\begin{split}
    m_2
    & =
    -
    \frac{1}{(2\pi)^n}
    \int_{B/\epsilon}
        \frac{1}{4}
        \frac{d^2}{d\epsilon^2}
        \Big(
            L^T
            \mathbb{D}_\epsilon^{-1}(k)
            L
        \Big)_{\epsilon=0}
        \, \hat{f}(k) \cdot \hat{f}^*(k)
    \, dk ,
\end{split}
\end{equation}
which is the second-order correction to the energy of the metamaterial at equilibrium. In the particular case of metamaterials with one single type of joint, $N=1$, $L$ reduces to the identity and (\ref{m2N}) simplifies to
\begin{equation}\label{m2N1}
\begin{split}
    m_2
    & =
    \frac{1}{(2\pi)^n}
    \int_{B/\epsilon}
        \frac{1}{4}
            \mathbb{D}''_0(k)
        \, \hat{u}_0(k) \cdot \hat{u}_0^*(k)
    \, dk ,
\end{split}
\end{equation}
which gives the second-order energy correction directly in terms of the zeroth-order displacements $u_0(x)$.

A further analysis shows that $\mathbb{D}''_0(k)
        \, \hat{u}_0(k) \cdot \hat{u}_0^*(k)$ is homogeneous of degree $4$ with regard to deflections and of degree $2$ with regard to rotations, whereupon an inverse Fourier transform gives
\begin{equation} \label{D1cmbX}
    m_2
    =
    \int_\Omega
        W_{2}(D\theta_0(x), D^2v_0(x))
    \, dx ,
\end{equation}
for some quadratic function $W_2$, where we have further localized the functional to the domain $\Omega$. Thus, the second-order energy correction can be expressed directly in terms of the gradients and second gradients of the zeroth-order rotations and deflections, respectively.

\subsection{The octet metamaterial}
\begin{figure*}[h!]
\begin{center}
    \subfigure[]{\includegraphics[width=0.8\textwidth]{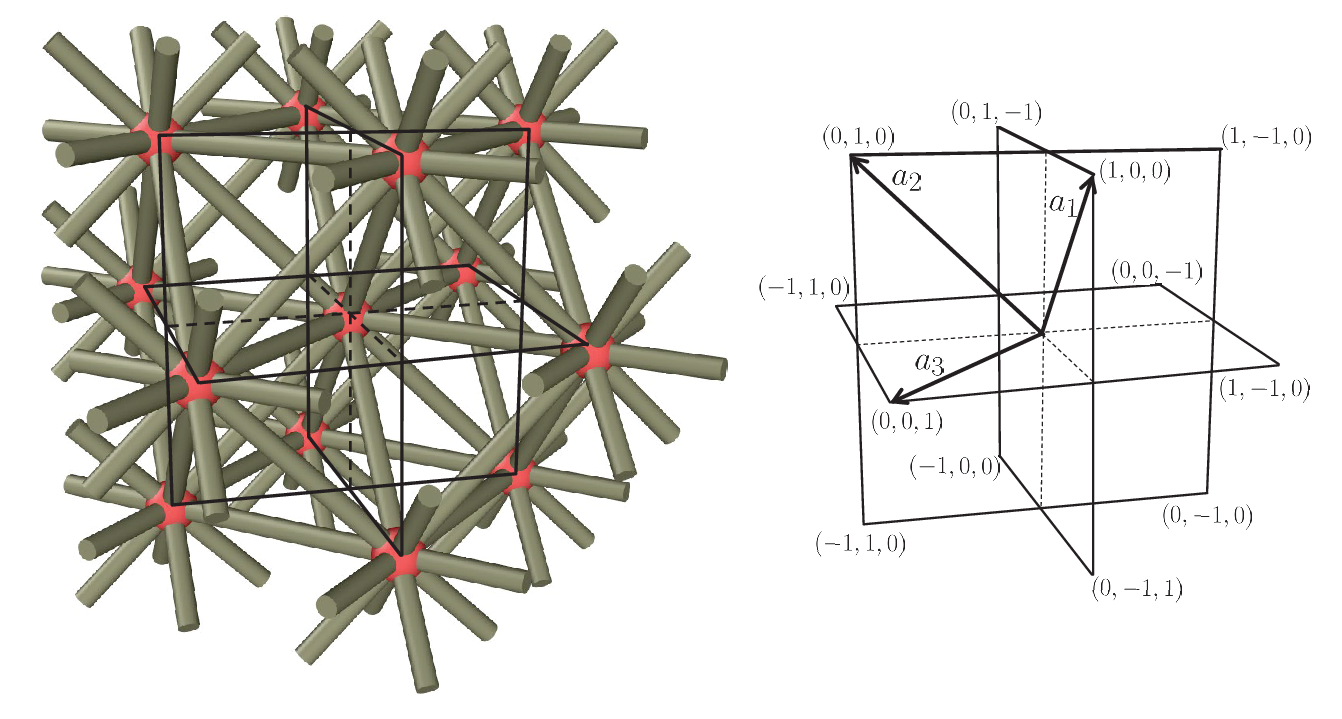}}
    \subfigure[]{\includegraphics[width=0.7\textwidth]{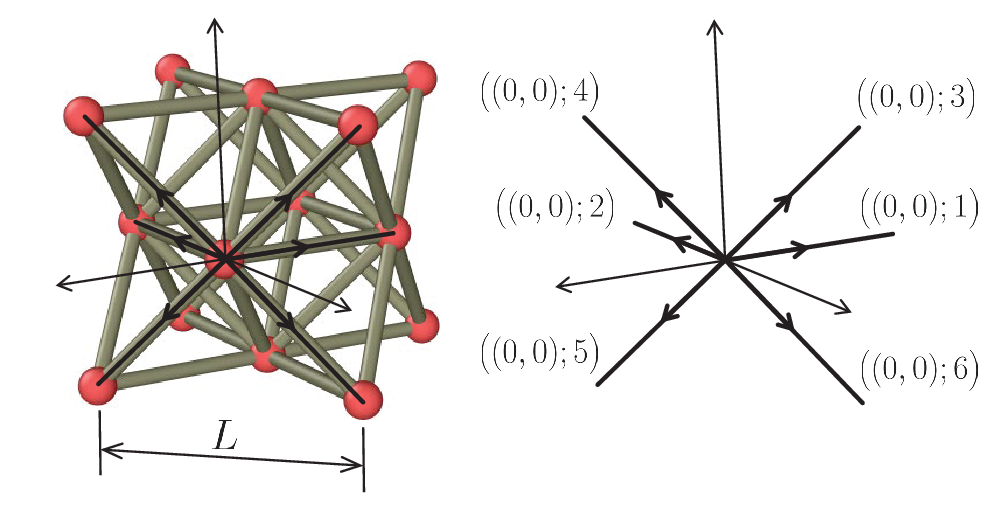}}
    \caption{Octet-truss metamaterial. (a) Joint numbering scheme using a simple Bravais lattice. (b) Bar numbering scheme.} \label{S2rFkk}
\end{center}
\end{figure*}

The octet-truss metamaterial of size $L$, Fig.~\ref{S2rFkk}, is three dimensional, $n=3$, and contains one type of joints, $N=1$, and six types of oriented bars, $M=6$, of length $l = L/\sqrt{2}$.

\subsubsection{Zeroth-order homogenization limit}
Assuming, for simplicity, $GI_1 := GJ$, $EI_2 = EI_3 := EI$, a direct computation of the limit (\ref{lIKdde}) gives, to zeroth order, the limiting continuum energy \cite{Ariza:2024}
\begin{equation} \label{f5CtCF}
\begin{split}
    E_0(u_0)
    & =
    \int_\Omega
        \frac{1}{2}
        \mathbb{C} \,
        \varepsilon_0(x)
        \cdot
        \varepsilon_0(x)
    \, dx
    \\ & +
    \int_\Omega
        \frac{1}{2}
        \frac{48 \sqrt{2} {EI}}{L^4}
        \big| \theta_0(x) - \frac{1}{2} \operatorname{curl} v_0(x) \big|^2
    \, dx ,
\end{split}
\end{equation}
comprising an elastic and a micropolar component. The
effective elastic moduli defining the former have cubic symmetry and evaluate to \cite{Ariza:2024}
\begin{equation} \label{s2SThn}
\begin{split}
    &
    C_{11}
    =
    \frac{4 {EA} L^2+24 {EI}}{\sqrt{2} L^4} ,
    \\ &
    C_{12}
    =
    \frac{\sqrt{2} \left({EA} L^2-6 {EI}\right)}{L^4} ,
    \\ &
    C_{44}
    =
    \frac{2 {EA} L^2+12 {EI}}{\sqrt{2} L^4} .
\end{split}
\end{equation}
For the octet truss, $EI=0$, these moduli match those reported by \cite{deshpande2001b}. The micropolar energy density is isotropic, as expected from the cubic symmetry of the lattice. It is also interesting to note that the effective moduli are independent of the torsional stiffness $GJ$ of the bars.

\subsubsection{Full three-dimensional second-order correction}
The first-order correction to the energy vanishes by the centrosymmetry of the lattice. To second order, a direct evaluation of (\ref{m2N1}) using the symbolic calculus code {\tt Mathematica} (Wolfram Research, Inc.) gives the second-order energy density in (\ref{D1cmbX}) as
\begin{equation}
    W_2
    =
    W_{2,EA} + W_{2,GJ} +
    W_{2,EI,11} + W_{2,EI,12} + W_{2,EI,21} + W_{2,EI,22} ,
\end{equation}
with
\begin{equation}
\begin{split}
    &
    W_{2,EA}(D\theta_0(x), D^2v_0(x))
    =
    \\ & -
    \frac{EA}{24 \sqrt{2}}
    \Big(
        6 (v_{1,12}^2 + v_{1,13}^2 )
        +
        2 v_{1,11}^2 + v_{1,22}^2 + v_{1,33}^2
   \Big)
    \\ & -
    \frac{\sqrt{2} EA}{12}
    v_{1,12} (v_{2,11} + v_{2,22})
    -
    \frac{\sqrt{2} EA}{12}
    v_{1,13} (v_{3,11} + v_{3,33})
    \\ & +
    \text{permutations} \;\,
    \{1,2,3\}\to\{2,3,1\}
    \;\,\text{and}\;\,
    \{2,3,1\}\to\{3,1,2\}
\end{split}
\end{equation}
as the axial contribution,
\begin{equation}
\begin{split}
    &
    W_{2,GJ}(D\theta_0(x), D^2v_0(x))
    = \\ &
    \frac{\sqrt{2} {GJ}}{2 L^2}
    \left(2 \theta_{1,1}^2 +\theta_{1,2}^2+\theta_{1,3}^2\right)
    +
    \frac{\sqrt{2} {GJ}}{L^2} \theta_{1,1} \theta_{2,2}
    +
    \frac{\sqrt{2} {GJ}}{L^2} \theta_{1,1} \theta_{3,3}
    \\ & +
    \text{permutations} \;\,
    \{1,2,3\}\to\{2,3,1\}
    \;\,\text{and}\;\,
    \{2,3,1\}\to\{3,1,2\}
\end{split}
\end{equation}
as the torsional contribution,
\begin{equation}
    W_{2,EI,11}(D\theta_0(x), D^2v_0(x)) =
    \qquad\qquad\qquad\qquad\qquad\qquad\qquad
\end{equation}
\begin{equation} \nonumber
\begin{split}
    & -
    \frac{{EI}}{4\sqrt{2} L^2}
    \Big(
        2 v_{1,11}^2
        +
        6 v_{1,11}  (v_{1,22} + v_{1,33})
        +
        3 (4 v_{1,22} v_{1,33}+v^2_{1,22}+v^2_{1,33})
    \Big)
    \\ & +
    \frac{\sqrt{2} {EI} }{2 L^2}
    v_{1,12} (v_{2,11}+v_{2,22})
    +
    \frac{\sqrt{2} {EI} }{2 L^2}
    v_{1,13} (v_{3,11}+v_{3,33})
    \\ & +
    \text{permutations} \;\,
    \{1,2,3\}\to\{2,3,1\}
    \;\,\text{and}\;\,
    \{2,3,1\}\to\{3,1,2\}
\end{split}
\end{equation}
as the deflection term in the bending contribution,
\begin{equation}
    W_{2,EI,12}(D\theta_0(x), D^2v_0(x))
    +
    W_{2,EI,21}(D\theta_0(x), D^2v_0(x))
    =
    \qquad
\end{equation}
\begin{equation}\nonumber
\begin{split}
    &
    \frac{\sqrt{2} {EI} }{L^2}
    (3 (v_{1,11}+v_{1,22})+2 v_{1,33})
    \, \theta_{2,3}
    -
    \frac{\sqrt{2} {EI} }{L^2}
    (3 (v_{1,11}+v_{1,33})+2 v_{1,22})
    \, \theta_{3,2}
    \\ & +
    \text{permutations} \;\,
    \{1,2,3\}\to\{2,3,1\}
    \;\,\text{and}\;\,
    \{2,3,1\}\to\{3,1,2\}
\end{split}
\end{equation}
as the coupling deflection-rotation term contribution, and
\begin{equation}
\begin{split}
    &
    W_{2,EI,22}(D\theta_0(x), D^2v_0(x))
    =
    \\ &
    -
    \frac{\sqrt{2} {EI} }{4 L^2}
    \left(
        2 \theta^2_{1,1}
        +
        3 \left(\theta^2_{1,2}+\theta^2_{1,3}\right)
    \right)
    +
    \frac{\sqrt{2} {EI} }{2 L^2} \theta_{1,1} \theta_{2,2}
    +
    \frac{\sqrt{2} {EI} }{2 L^2} \theta_{1,1} \theta_{3,3}
    \\ & +
    \text{permutations} \;\,
    \{1,2,3\}\to\{2,3,1\}
    \;\,\text{and}\;\,
    \{2,3,1\}\to\{3,1,2\}
\end{split}
\end{equation}
as the rotation term in the bending contribution.

\subsubsection{Second-order correction in plane strain}
Under {\sl plane-strain conditions}, the above energy corrections reduce to
\begin{equation}
\begin{split}
    &
    W_{2,EA}(D\theta_0(x), D^2v_0(x))
    =
    \\ & -
    \frac{EA}{24 \sqrt{2}}
    \Big(
        6 v_{1,12}^2
        +
        2 v_{1,11}^2 + v_{1,22}^2
    \Big)
    -
    \frac{\sqrt{2} EA}{12}
    v_{1,12} (v_{2,11} + v_{2,22})
    \\ & -
    \frac{\sqrt{2} EA}{12}
    v_{2,12} (v_{1,11} + v_{1,22})
    -
    \frac{EA}{24 \sqrt{2}}
    \Big(
        6 v_{2,12}^2
        +
        2 v_{2,11}^2 + v_{2,22}^2
    \Big)
\end{split}
\end{equation}
as the axial contribution,
\begin{equation}
\begin{split}
    &
    W_{2,GJ}(D\theta_0(x), D^2v_0(x))
    =
    \frac{\sqrt{2} {GJ}}{2 L^2}
    \left(\theta,_1^2 + \theta,_2^2\right)
\end{split}
\end{equation}
as the torsional contribution,
\begin{equation}
    W_{2,EI,11}(D\theta_0(x), D^2v_0(x)) =
    \qquad\qquad\qquad\qquad\qquad\qquad\qquad
\end{equation}
\begin{equation} \nonumber
\begin{split}
    & -
    \frac{{EI}}{4\sqrt{2} L^2}
    \Big(
        2 v_{1,11}^2
        +
        6 v_{1,11} v_{1,22}
        +
        3 v_{1,22}^2
    \Big)
    +
    \frac{\sqrt{2} {EI} }{2 L^2}
    v_{1,12} (v_{2,11}+v_{2,22})
    \\ & +
    \frac{\sqrt{2} {EI} }{2 L^2}
    v_{2,12} (v_{1,11}+v_{1,22})
    -
    \frac{{EI}}{4\sqrt{2} L^2}
    \Big(
        2 v_{2,11}^2
        +
        6 v_{2,11} v_{2,22}
        +
        3 v_{2,22}^2
    \Big)
\end{split}
\end{equation}
as the deflection term in the bending contribution,
\begin{equation}
    W_{2,EI,12}(D\theta_0(x), D^2v_0(x))
    +
    W_{2,EI,21}(D\theta_0(x), D^2v_0(x))
    =
    \qquad
\end{equation}
\begin{equation}\nonumber
\begin{split}
    &
    -
    \frac{\sqrt{2} {EI} }{L^2}
    (3 v_{1,11}+2 v_{1,22})
    \, \theta,_2
    +
    \frac{\sqrt{2} {EI} }{L^2}
    (3 v_{2,11}+2 v_{2,22})
    \, \theta,_1
\end{split}
\end{equation}
as the coupling deflection-rotation term contribution, and
\begin{equation}
\begin{split}
    &
    W_{2,EI,22}(D\theta_0(x), D^2v_0(x))
    =
    -
    \frac{3 EI}{\sqrt{2} L^2}
    ( \theta,_1^2 + \theta,_2^2 )
\end{split}
\end{equation}
as the rotation term in the bending contribution.

\subsubsection{Second-order correction in plane strain with relaxed rotations}

A further simplification is accrued if the zeroth-order joint rotations $\theta_0(x)$ are {\sl relaxed}, i.~e., if
\begin{equation}
    \theta_0(x) = \frac{1}{2} \operatorname{curl} v_0(x) ,
\end{equation}
as may be expected if no distributed torques are applied and the joint rotations are left free at the boundary. Under these conditions, the various terms in the second-order correction reduce to
\begin{equation} \label{K3pWms}
\begin{split}
    &
    W_{2,EA}(D^2v_0(x)) =
    \\ & -
    \frac{EA}{24\sqrt{2}}
    (2 v^2_{1,11} + 6 v^2_{1,12} + v^2_{1,22})
    -
    \frac{EA}{6\sqrt{2}}
    v_{1,12}(v_{2,11}+v_{2,22})
    \\ & -
    \frac{EA}{6\sqrt{2}}
    v_{2,12}(v_{1,11}+v_{1,22})
    -
    \frac{EA}{24\sqrt{2}}
    (2 v^2_{2,22} + 6 v^2_{2,12} + v^2_{2,11})
\end{split}
\end{equation}
for the axial contribution,
\begin{equation} \label{SgY6bf}
\begin{split}
    &
    W_{2,GJ}(D^2v_0(x)) =
    \\ &
    \frac{GJ}{4\sqrt{2} L^2}
    (v^2_{1,22} + v_{1,11} v_{1,22} )
    -
    \frac{GJ}{4\sqrt{2} L^2}
    v_{1,12} (v_{2,11} + v_{2,22})
    - \\ &
    \frac{GJ}{4\sqrt{2} L^2}
    v_{2,12} (v_{1,11} + v_{1,22})
    +
    \frac{GJ}{4\sqrt{2} L^2}
    (v^2_{2,11} + v_{2,11} v_{2,22} )
\end{split}
\end{equation}
for the torsional contribution, and
\begin{equation} \label{55rMK6}
\begin{split}
    &
    W_{2,EI}(D^2v_0(x)) =
    \\ & -
    \frac{EI}{8\sqrt{2}L^2}
    (4 v^2_{1,11} - 9 v^2_{1,12} - 7 v^2_{1,22})
    -
    \frac{9 EI}{8\sqrt{2}L^2}
    v_{1,12}(v_{2,11}+v_{2,22})
    \\ & -
    \frac{9 EI}{8\sqrt{2}L^2}
    v_{2,12}(v_{1,11}+v_{1,22})
    -
    \frac{EI}{8\sqrt{2}L^2}
    (4 v^2_{2,22} - 9 v^2_{2,12} - 7 v^2_{2,11})
\end{split}
\end{equation}
for the bending contribution.

\subsection{Example: Octet metamaterial in uniaxial strain}

\begin{figure}[ht!]
\begin{center}
    \includegraphics[width=0.85\textwidth]{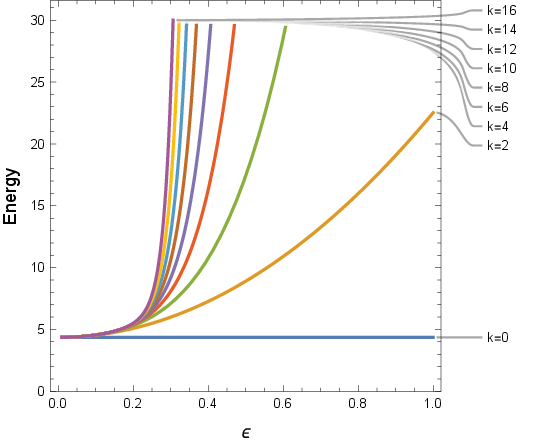}
    \caption{Dipole in octet metamaterial constrained to deform in uniaxial strain. $\Gamma$-expansions of energy from zeroth-order to sixteenth-order.} \label{8KG8zf}
\end{center}
\end{figure}

We illustrate the convergence properties of $\Gamma$-expansion by means of a simple example concerned with an octet metamaterial constrained to deform in uniaxial strain in the direction $x_3$ and loaded by a force dipole. In this case, the scaled dynamical matrix reduces to
\begin{equation}
    \mathbb{D}_\epsilon(k)
    =
    \frac
    {
        8 \sqrt{2}
        \left({EA} L^2+6 {EI}\right)
    }
    {
        {\epsilon}^2 L^6
    }
    \left(
        1
        -
        \cos \frac{{\epsilon} {k_3} L}{ \sqrt{2}}
    \right).
\end{equation}
Considering a dipole of strength $F$ and width $b$,
\begin{equation}
    \hat{f}_3(k_3)
    =
    -2 i F {k_3} {\rm e}^{-\frac{1}{4} b^2 {k_3}^2},
\end{equation}
the minimum potential energy (\ref{meps}) reads
\begin{equation}
\begin{split}
    m_\epsilon
    & =
    -
    \frac{1}{2\pi}
    \int_{-\infty}^{\infty}
        \frac{1}{2}
        \frac{|\hat{f}_3(k_3)|^2}{\mathbb{D}_\epsilon(k_3)}
    \, dk_3
    \\ & =
    -
    \frac{1}{2\pi}
    \int_{-\infty}^{\infty}
        \frac
        {
            {\epsilon}^2 F^2 {k_3}^2 L^6
        }
        {
            \sqrt{2} \left(8 {EA} L^2+48 {EI}\right)
        }
        {\rm e}^{-\frac{1}{2} b^2 k_3^2}
        \csc^2 \frac{{\epsilon} {k_3} L}{2 \sqrt{2}}
    \, dk_3 .
\end{split}
\end{equation}
For fixed $F$ and $b$, the expansion (\ref{bzynWH}) at eight-order gives
\begin{equation}
\begin{split}
    m_\epsilon
    & \sim
    \frac{F^2 L^4}{2 \sqrt{\pi } \left(b {EA} L^2+6 b {EI}\right)}
    +
    \frac{\epsilon^2 F^2 L^6}{48 \sqrt{\pi } b^3 \left({EA} L^2+6 {EI}\right)}
    \\ &+
    \frac{\epsilon^4 F^2 L^8}{640 \sqrt{\pi } b^5 \left({EA} L^2+6 {EI}\right)}
    +
    \frac{5 \epsilon^6 F^2 L^{10}}{32256\sqrt{\pi } b^7 \left({EA} L^2+6{EI}\right)}
    \\ &+
    \frac{7 \epsilon^8 F^2 L^{12}}{368640 \sqrt{\pi } b^9 \left({EA}L^2+6 {EI}\right)}
    +
    O(\epsilon^{10}).
\end{split}
\end{equation}
The normalized energy is shown in Fig.~\ref{8KG8zf} to various orders. We see that the $\Gamma$-expansion of order $k$ exhibits $k$-th order convergence to the zeroth-order homogenized energy as $\epsilon\to 0$, as required. We also observe that, for fixed $\epsilon$, the $\Gamma$-expansion converges to the exact energy as the degree of approximation $k$ increases.

\section{Application to fracture of metamaterials}

Suppose that the base material is perfectly brittle and is characterized by a critical energy-release rate $G_c$, or specific energy per unit area. Further suppose that crack growth is planar on average and that the ratio of base-material fractured area to nominal fractured area scales with the relative density $\bar{\rho}$, i.~e., the ratio of the density of the lattice material to the density of the base material from which it is made. To first order and neglecting the effect of double counting at the joints, the relative density bears the relation
\begin{equation} \label{UCgR4X}
    \bar{\rho} = c \frac{A}{L^2} ,
\end{equation}
where the constant $c$ is characteristic of the geometry of the metamaterial and $L$ is the size of the unit cell. For an octet structure \cite{deshpande2001b} and with the choice of $L$ shown in Fig.~\ref{S2rFkk},
\begin{equation} \label{Y9zSnV}
    c = 12 \sqrt{2}.
\end{equation}
Under these assumptions, the fracture energy expended per unit area of crack advance is $\bar{\rho} G_c$, and an appeal to Griffith's criterion (see, e.~g., \cite{Hutchinson:1979}) requires that
\begin{equation} \label{c848rN}
    G = \bar{\rho} G_c
\end{equation}
at crack-growth initiation, where $G$ is the energy-release rate per unit crack advance in the metamaterial.

We remark that the simple Griffith criterion (\ref{c848rN}) used here is intended to facilitate the use of variational energy-based methods such as developed in the foregoing. Other failure criteria have been proposed and extensively investigated based on a variety of assumptions, including failure stresses and strains for the base material of the bars under tensile and bending conditions (\cite{Gibson:1999,Fleck:2007,Fleck2010}). The simple Griffith criterion (\ref{c848rN}) applies to metamaterials under axial-dominated conditions. However, these alternative failure criteria, while affecting the right-hand side of the crack-growth initiation criterion (\ref{c848rN}), have no effect on the driving force $G$ for fracture and its dependence on configuration and unit-cell size, which is the primary focus of the analysis that follows, and do not alter its main conclusions.

An elucidation of the effect of metamaterial structure and size on crack-growth initiation therefore requires a determination of the energy release rate $G$ as a function of the particular domain, loading, and crack size and geometry under consideration. By way of reference, in Section~\ref{4TcJ7F} we begin by considering the energy-release rate $G_0$ predicted in the zeroth-order homogenization limit, which is fully characterized by the effective elastic moduli of the homogenized solid and the stress-intensity factor of the elastic $K$-field attendant to the crack tip. Thus, the homogenized solid exhibits linear-elastic fracture mechanics scaling and, in particular, the effective toughness of the metamaterial, as set forth by (\ref{c848rN}), is independent of the unit-cell size.

In order to capture the experimentally-observed size effect \cite{Shaikeea2022}, in Section~\ref{W4e6hM} we endeavor to retain mesoscale information by extending the homogenization analysis to second order. In this case, the predicted energy release rate $G_2$ and apparent toughness of the metamaterial depend on the specific configuration of the specimen and loading, including unit-cell size. In order to characterize such dependence analytically, we consider a simple double-cantilever beam configuration and resort to compliance analysis to determine the requisite energy release rate (see, e.~g., \cite{Hutchinson:1979} for background).

\subsection{Zeroth-order homogenization} \label{4TcJ7F}

To zeroth order, the homogenized metamaterial is indistinguishable from an elastic solid characterized by its effective moduli, irrespective of the size and shape of the domain and the loading. In this limit, the elastic solution for a cracked solid exhibits a $K$-field asymptotically near the crack tip. For orthotropic materials deforming in mode I with the crack directions aligned with the directions of material symmetry, the energy-release rate is related to the stress-intensity factor $K_0$ as \cite{BankSills:2005}
\begin{equation} \label{peHt4C}
    G_0
    =
    \frac{K_0^2}{E_0'}
\end{equation}
where
\begin{equation} \label{Q4Fn8b}
    \frac{1}{E_0'}
    =
    \frac{1}{2}
    \Big(
        2 \sqrt{S_{11}S_{22}} + 2 S_{12} + S_{44}
    \Big)^{1/2}
    \sqrt{S_{22}} ,
\end{equation}
$E_0'$ is a plane-strain Young's modulus and $S_{ij}$ are reduced plane-strain elastic compliances. For a cubic material, (\ref{Q4Fn8b}) specializes to
\begin{equation}
    S_{11}
    =
    S_{22}
    =
    \frac{C_{11}}{C_{11}^2-C_{12}^2} ,
    \quad
    S_{12}
    =
    \frac{C_{12}}{C_{12}^2-C_{11}^2} ,
    \quad
    S_{44}
    =
    \frac{1}{C_{44}} ,
\end{equation}
where we assume $C_{12} > 0$, $C_{11} > C_{12}$, $C_{44} > 0$,
and (\ref{Q4Fn8b}) reduces to
\begin{equation} \label{Vjm69U}
    \frac{1}{E_0'}
    =
    \frac{1}{2}
    \sqrt
    {
        \frac
        {
            C_{11} (C_{11}+C_{12}+2 C_{44})
        }
        {
            C_{44} (C_{11}-C_{12}) (C_{11}+C_{12})^2
        }
    } .
\end{equation}
For isotropic materials, (\ref{Vjm69U}) further specializes to
\begin{equation}
    E_0' = \frac{E}{1-\nu^2} ,
\end{equation}
with $E$ and $\nu$ the {\sl effective} Young's modulus and Poisson's ratio \cite{Rice:1968}.
Inserting (\ref{peHt4C}) into (\ref{c848rN}), we obtain
\begin{equation} \label{3HZxxR}
    K_{c,0} = \sqrt{E_0' \, \bar{\rho} G_c} ,
\end{equation}
which relates the critical stress-intensity factor, or toughness, of the homogenized metamaterial to its elastic moduli, relative density and the critical energy-release rate of the base material.

It should be carefully noted that $E_0'$ in (\ref{3HZxxR}) itself depends on the structure of the metamaterial. By way of illustration, we elucidate this dependence for the octet frame. In this case, inserting (\ref{s2SThn}) into (\ref{Vjm69U}) gives
\begin{equation} \label{7WrSvd}
    \frac{1}{E_0'}
    =
    \frac{f(\lambda)}{\bar{\rho}  E},
    \quad
    f(\lambda)
    =
    \frac
    {
        \lambda ^2
    }
    {
        \lambda^2+2
    }
    \sqrt
    {
        10
        -
        \dfrac{144}{\lambda ^2+18},
    }
\end{equation}
where we have used (\ref{UCgR4X}) and (\ref{Y9zSnV}) and we have introduced the slenderness ratio of the bars
\begin{equation}\label{slenderness}
    \lambda = \frac{l}{r},
    \quad
    r = \sqrt{\frac{I}{A}} ,
\end{equation}
in terms of the radius of gyration $r$ of the cross-section and the length $l$ of the bar. For the octet truss, corresponding to the limit $\lambda\to +\infty$, (\ref{7WrSvd}) further specializes to
\begin{equation} \label{6cFdCg}
    \frac{1}{E_0'}
    =
    \frac{\sqrt{10}}{\bar{\rho} E} .
\end{equation}
Inserting (\ref{7WrSvd}) and (\ref{6cFdCg}) into (\ref{3HZxxR}), we finally obtain
\begin{equation} \label{vsJ7gC}
    K_{c,0} = \bar{\rho} \, \sqrt{\frac{ E G_c}{f(\lambda)} } ,
    \quad
    K_{c,0} = \bar{\rho} \, \sqrt{\frac{ E G_c}{\sqrt{10}} } ,
\end{equation}
for the octet frame and truss, respectively.

\begin{figure}[ht!]
\begin{center}
    \includegraphics[width=0.75\textwidth]{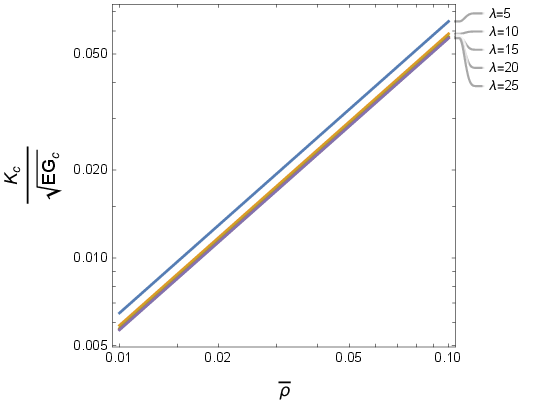}
    \caption{Zeroth-order homogenized octet frame. Dependence of normalized mode-I toughness $K_{c,0}/\sqrt{E G_c}$ on relative density $\bar{\rho}$ and slenderness $\lambda$ of the bars.} \label{UmE5nY}
\end{center}
\end{figure}

The dependence of the normalized toughness $K_{c,0}/\sqrt{E G_c}$ predicted by (\ref{vsJ7gC}) on the relative density $\bar{\rho}$ and the slenderness $\lambda$ of the bars is shown in Fig.~\ref{UmE5nY}. Note from~\eqref{UCgR4X} and~\eqref{slenderness} that $\lambda$ may be varied independently of $\bar{\rho}$ by considering, at fixed bar length $l$, cross-sections with varying moment of inertia $I$ but equal cross-sectional area $A$. We note that, to zeroth order, the toughness $K_c$ of the octet scales linearly with the relative density $\bar{\rho}$, in agreement with \cite{Fleck:2007, Fleck2010}, but is otherwise independent of the shape or size of the domain and crack, the loading and the unit-cell size. Thus, as expected, zeroth-order homogenization wipes out all mesoscopic information about the metastructure and, in particular, is insufficient to predict size effects such as reported in \cite{Shaikeea2022}.

It is also noteworthy that the effective toughness of metamaterial frames depends on the slenderness of the bars, which effectively measures the ratio of the axial to the bending stiffness of the bars. In particular, the effect of bending is non-negligible for small $\lambda$ and becomes increasingly less important as $\lambda$ increases, with articulated truss structures corresponding to the limit of $\lambda\to +\infty$ (cf.~\cite{deshpande2001a} for a classification of metamaterials as stretching and bending dominated). We note from Fig.~\ref{UmE5nY} that the effect of bending is to increase the effective toughness of metamaterial frames with respect to that of the corresponding metamaterial trusses.

\subsection{Second-order homogenization} \label{W4e6hM}

\begin{figure}[ht!]
\begin{center}
    \includegraphics[width=0.99\textwidth]{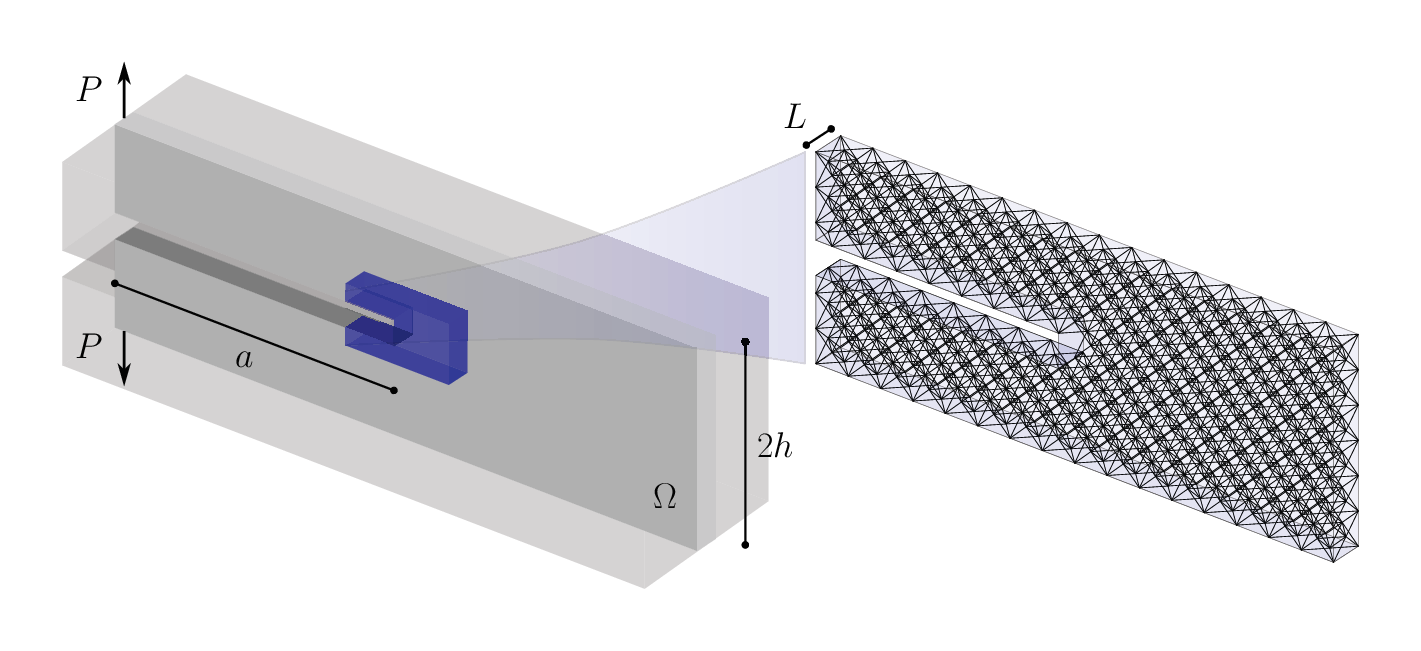}
    \caption{Octet-frame metastructure, double-cantilever beam configuration.} \label{8Bmv9c}
\end{center}
\end{figure}

As already mentioned, the energy-release rate $G$ in (\ref{c848rN}) is a structural configurational force that depends on the particular size and geometry of the solid and crack, the loading and, at the mesoscale, on the unit-cell size. We aim to retain sufficient mesoscale information to elucidate size effects by extending the homogenization analysis to second order.

For definiteness, we consider an ideal specimen made of octet metamaterial in the form of a semi-infinite plate of height $2h$ containing a crack of length $a$ within its midplane abutting on the edge of the plate, Fig.~\ref{8Bmv9c}. The plate deforms elastically under the action of wedging tip loads per unit length $P$. We estimate the attendant energy-release rate $G$ by recourse to plate theory and compliance analysis \cite{Hutchinson:1979}.

\subsubsection{Compliance analysis} \label{7GHpCL}
We briefly recall the fundamentals of compliance analysis (see, e.~g., \cite{Hutchinson:1979} for further details). Suppose that a prismatic linear elastic solid deforms in plane strain under the action of loads imparted by means of a loading device such that
\begin{equation}
    \int_{\partial\Omega} t_i u_i  dS = P \Delta ,
\end{equation}
where $P$ and $\Delta$ are a generalized force per unit thickness and a conjugate generalized displacement, respectively. For a linear elastic solid containing a crack of length $a$, at equilibrium $P$ and $\Delta$ necessarily bear a linear relation of the form
\begin{equation}
    \Delta = C P ,
\end{equation}
where $C$ is the {\sl compliance} of the cracked solid. Suppose that the loading device is soft, resulting in force control. By Clapeyron's theorem, at equilibrium the potential energy attains the value
\begin{equation}
    {\Pi}
    =
    \frac{1}{2} P \Delta - P \Delta
    =
    -
    \frac{C}{2}  P^2 ,
\end{equation}
and the corresponding energy-release rate follows as
\begin{equation} \label{xG8MHv}
    G
    =
    -
    \left.
    \frac{\partial{\Pi}}{\partial a}
    \right|_P
    =
    \frac{P^2}{2} \frac{dC}{da} .
\end{equation}
A similar analysis for the stiff load device gives the same result \cite{Hutchinson:1979}. In cases where the compliance is known or can be estimated, identity (\ref{xG8MHv}) provides a shortcut for the evaluation of the energy-release rate.

\subsubsection{Zeroth-order compliance analysis}
To zeroth order, using elementary plate theory as an approximation \cite{Timoshenko:1959}, the opening displacement follows as
\begin{equation} \label{wv24Vm}
    \Delta = \frac{2 P a^3}{3 D_0} ,
    \quad
    D_0 = \frac{E_0 h^3}{12},
\end{equation}
where $D_0$ is the bending stiffness of the plate and $E_0$ is the plane-strain Young's modulus of the metamaterial, which we assume to be orthotropic with principal directions aligned with the plate and the crack. For isotropic materials,
\begin{equation}
    E_0 = \frac{E}{1-\nu^2},
    \quad
    D_0 =\frac{E h^3}{12(1-\nu^2)} ,
\end{equation}
with $E$ and $\nu$ the {\sl effective} Young's modulus Poisson's ratio. From (\ref{xG8MHv}) and  (\ref{wv24Vm}), the compliance and the energy-release rate follow as
\begin{equation} \label{9qMSPB}
    C_0 = \frac{8 a^3}{E_0 h^3} ,
    \quad
    G_0
    =
    \frac{12 P^2 a^2}{E_0 h^3} ,
\end{equation}
respectively, where we have used (\ref{xG8MHv}) in the calculation of $G_0$.

For a material with cubic symmetry, the plane-strain Young's modulus follows as
\begin{equation} \label{47x9GB}
    E_0
    =
    C_{11}-\frac{C_{12}^2}{C_{11}} ,
\end{equation}
where $C_{11}$, $C_{12}$ and $C_{44}$ are the cubic elastic moduli of the metamaterial. For the octet-frame, eqs.~(\ref{f5CtCF}), the plane-strain Young's modulus (\ref{47x9GB}) evaluates to
\begin{equation} \label{kwSxu9}
    E_0
    =
    \frac
    {
        3 \left({EA} L^2+2 {EI}\right) \left({EA} L^2+18 {EI}\right)
    }
    {
        \sqrt{2} L^4 \left({EA} L^2+6{EI}\right)
    }
\end{equation}
which, inserted into (\ref{9qMSPB}) yields the compliance
\begin{equation} \label{4g2n5C}
    C_0
    =
    \frac{8 a^3}{h^3}
    \frac
    {
        \sqrt{2} L^4 \left({EA} L^2+6{EI}\right)
    }
    {
        3 \left({EA} L^2+2 {EI}\right) \left({EA} L^2+18 {EI}\right)
    } ,
\end{equation}
and the energy-release rate
\begin{equation} \label{nqSc4t}
    G_0
    =
    \frac{4 P^2 a^2}{h^3}
    \frac
    {
        \sqrt{2} L^4 \left({EA} L^2+6{EI}\right)
    }
    {
        \left({EA} L^2+2 {EI}\right) \left({EA} L^2+18 {EI}\right)
    } .
\end{equation}
For the octet-truss, corresponding to $EI=0$, the preceding relations simplify to
\begin{equation} \label{zZxA66}
    E_0
    =
    \frac{3 {EA}}{\sqrt{2} L^2} ,
    \quad
    C_0
    =
    \frac{8 \sqrt{2} a^3 L^2}{3 {EA} h^3} ,
    \quad
    G_0
    =
    \frac{4 \sqrt{2} a^2 L^2 P^2}{{EA} h^3} ,
\end{equation}
respectively.

\subsubsection{Second-order compliance analysis}
In order to retain unit-cell size information in the effective behavior of the homogenized metamaterial at the mesoscale, we extend the analysis to second order. From elementary plate theory \cite{Timoshenko:1959}, the displacement field follows as
\begin{equation} \label{zShx2G}
    u_2(x_1,x_2)
    =
    \frac{\Delta}{2}
    \Big(
        1
        -
        \frac{3 x_1}{2 a}
        +
        \frac{x_1^3}{2 a^3}
    \Big) ,
    \quad
    0 \leq x_1 \leq a ,
    \quad
    x_2 \geq 0 ,
\end{equation}
and
\begin{equation} \label{4pPVb7}
\begin{split}
    u_1(x_1,x_2)
    & =
    -
    u_{2,1}(x_1,x_2) \Big(x_2 - \frac{h}{2}\Big)
    \\ & =
    \frac{\Delta}{2a}
    \Big(\frac{3}{2}-\frac{3x_1^2}{2 a^2}\Big)
    \Big(x_2 - \frac{h}{2}\Big) ,
    \quad
    0 \leq x_2 \leq h .
\end{split}
\end{equation}
Inserting (\ref{zShx2G}) and (\ref{4pPVb7}) into (\ref{K3pWms}), (\ref{55rMK6}) and (\ref{SgY6bf}) gives the second-order strain-energy density as
\begin{subequations}
\begin{align}
    &
    W_{2,EA}
    =
    -
    \frac{3 {EA} \left((h-2 {x_2})^2+4
    {x_1}^2\right)}{64 \sqrt{2} a^6}
    {\Delta}^2 ,
    \\ &
    W_{2,EI}
    =
    -
    \frac{9 {EI} (h-5 {x_1}-2 {x_2}) (h+5
    {x_1}-2 {x_2})}{32 \sqrt{2} a^6 L^2}
    {\Delta}^2 ,
    \\ &
    W_{2,GJ}
    =
    \frac{9 {GJ} {x_1}^2}{8 \sqrt{2} a^6 L^2}
    {\Delta}^2 .
\end{align}
\end{subequations}
Integrating over the domain $[0,a]\times[0,h]$ of one beam, the corresponding energies follow as
\begin{subequations} \label{rH3uC3}
\begin{align}
    &
    m_{2,EA}
    =
    -
    \frac{{EA} \left(4 a^2 h+h^3\right)}{32 \sqrt{2} a^5}
    \, {\Delta}^2 ,
    \\ &
    m_{2,EI}
    =
    \frac{3 {EI} \left(25 a^2 h-h^3\right)}{16 \sqrt{2} a^5 L^2}
    \, {\Delta}^2  ,
    \\ &
    m_{2,GJ}
    =
    \frac{3 {GJ} h}{4 \sqrt{2} a^3 L^2}
    \, {\Delta}^2 .
\end{align}
\end{subequations}
Thus, to second order,
\begin{equation}
    P
    =
    \frac{\Delta}{C_0(a)}
    +
    \frac{\partial m_2}{\partial\Delta}
    :=
    \frac{\Delta}{C_2(a,L)} .
\end{equation}
Using (\ref{9qMSPB}), (\ref{kwSxu9}) and (\ref{rH3uC3}), we find
\begin{equation} \label{YP6UKq}
\begin{split}
    \frac{1}{C_2(a,L)}
    & =
    \frac{h^3}{8 a^3}
    \frac
    {
        3 \left({EA} L^2+2 {EI}\right) \left({EA} L^2+18 {EI}\right)
    }
    {
        \sqrt{2} L^4 \left({EA} L^2+6{EI}\right)
    }
    \\ & -
    \frac
    {
        h \left({EA} L^2 \left(4 a^2+h^2\right)
        -
        6 a^2 (25 {EI}+4 {GJ})+6 {EI} h^2\right)
    }
    {
        16 \sqrt{2} a^5 L^2
    } ,
\end{split}
\end{equation}
where we explicate the dependence of $C_2(a,L)$ on $a$ and $L$ in order to emphasize that it depends on the crack configuration and the size of the unit cell. Evidently, $C_2$ also depends on the geometry and size of the domain, which in this case is represented by $h$, but this dependence is omitted for brevity.

Finally, to second-order, the energy-release rate follows from (\ref{YP6UKq}) and (\ref{xG8MHv}) as
\begin{equation} \label{tKzeV2}
    G_2(a,L,P)
    =
    \frac{P^2}{2} \frac{dC_2(a,L)}{da} .
\end{equation}
For the octet-truss, corresponding to $EI=GJ=0$, the preceding expressions simplify to
\begin{subequations}\label{z4GuDB}
\begin{align}
    &   \label{fFW9yq}
    C_2(a,L)
    =
    -
    \frac
    {
        16 \sqrt{2} a^5 L^2
    }
    {
        {EA} h L^2 \left(4 a^2+h^2\right)-6 a^2 {EA} h^3
    } ,
    \\ &
    G_2(a,L,P)
    =
    -
    \frac{18 \sqrt{2} D_0^2 h L^2 \left(6 a^2
       \left(2 L^2-3 h^2\right)+5 h^2 L^2\right)}{{EA} \left(a L^2
       \left(4 a^2+h^2\right)-6 a^3 h^2\right)^2}
    \, {\Delta}^2 ,
\end{align}
\end{subequations}
respectively.

\subsection{Direct verification from discrete bar calculations}

\begin{figure}[ht!]
\begin{center}
    \subfigure[]{\includegraphics[width=0.47\textwidth]{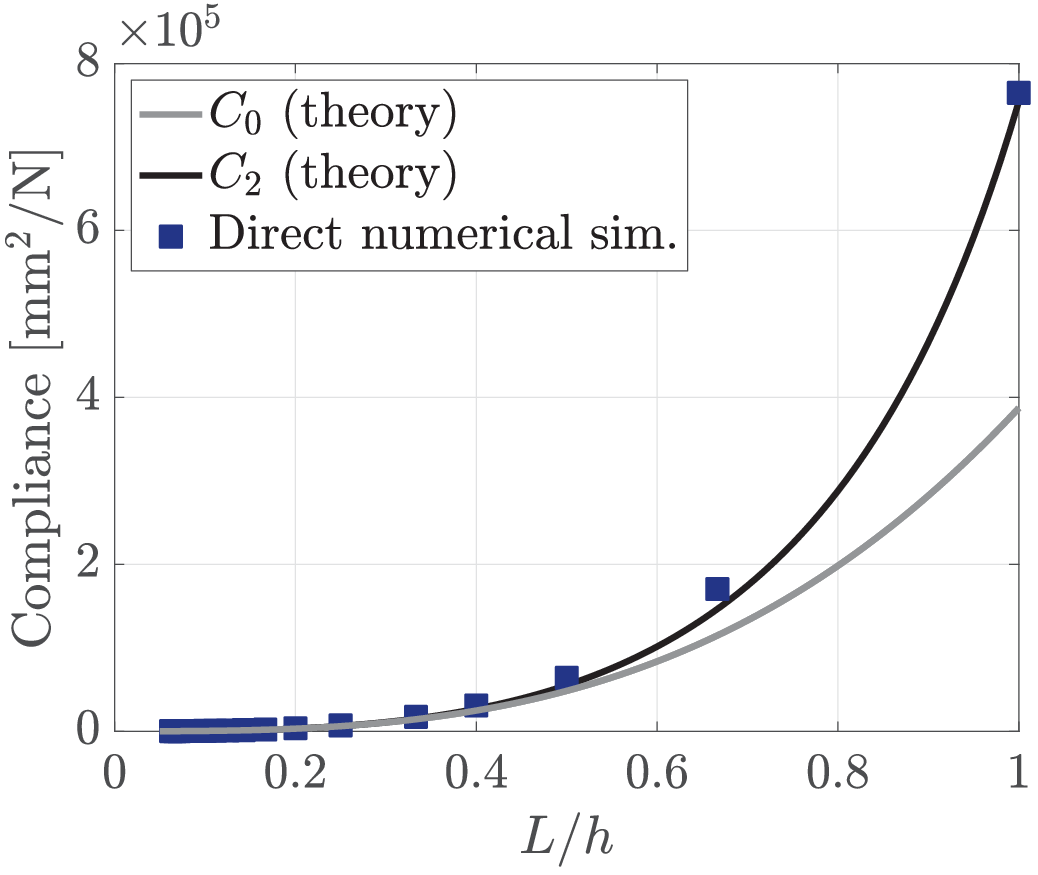}}
    \subfigure[]{\includegraphics[width=0.49\textwidth]{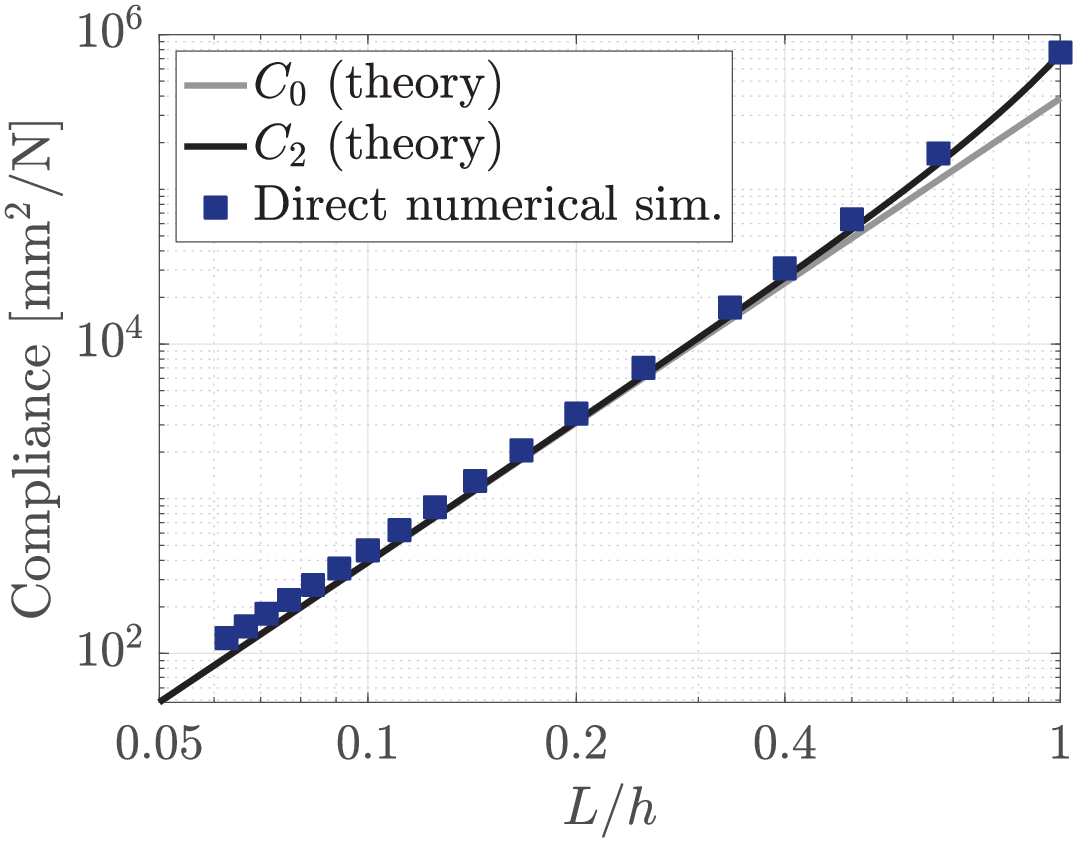}}
    \caption{Octet-frame metastructure, $EA = 1.43$ N, $EI = 3.77\times10^{-4}$ N mm$^2$, $L=0.75$ mm. Double-cantilever plate configuration, $a=48$ mm, $h\in[0.75,12]$. Compliance as a function ratio $L/h$ of cell size to plate thickness. Comparison of direct calculation using a bar model, zeroth-order continuum limit (\ref{4g2n5C}) and second-order continuum limit (\ref{YP6UKq}). (a)~Linear plot; (b) log plot. } \label{cddyD9}
\end{center}
\end{figure}

We aim to verify and assess the accuracy of the second-order formula (\ref{YP6UKq}) for the compliance, which is the basis for compliance analysis. To this end, we carry out direct numerical simulations of octet-frame structures in the double-cantilever plate configuration, Fig.~\ref{8Bmv9c}, $EA = 1.43$, $EI = 3.77\times10^{-4}$, $a=48$ and $h\in[0.75,12]$. The results are compared in Fig.~\ref{cddyD9} with the predictions of the zeroth-order and second-order formulae (\ref{4g2n5C}) and (\ref{YP6UKq}), respectively, over a range of unit-cell sizes. The improvement afforded by the second-order formula (\ref{YP6UKq}) over the zeroth-order formula (\ref{4g2n5C}) and the excellent accuracy afforded by the latter are evident from the figure.

\subsection{Size effect}

As surmised, the second-order relations retain mesoscale information and, as a result, they exhibit size effects. Thus, for fixed $a$ the critical load $P_c$ for crack-growth initiation follows from the Griffith condition
\begin{equation} \label{h8ghUb}
    G(a,L,P_c) = \bar{\rho} G_c .
\end{equation}
To second-order accuracy in metamaterial lattice size, we may replace (\ref{h8ghUb}) by
\begin{equation} \label{m8s7PS}
    G_2(a,L,P_c) \sim \bar{\rho} G_c ,
\end{equation}
with $G_2(a,L,P)$ given by the second-order analysis, e.~g., (\ref{tKzeV2}) for the octet-frame. Alternatively, since $G_0(a,L,P)$ and $G_2(a,L,P)$ are quadratic in $P$, (\ref{m8s7PS}) can be equivalently rewritten as
\begin{equation}
    G_0(a,P_c)
    \sim
    g^2(a,L) \bar{\rho} G_c ,
\end{equation}
where, from (\ref{xG8MHv}),
\begin{equation} \label{q2BzAw}
    g(a,L)
    =
    \Big(
        \frac{\partial C_0(a)/\partial a}{\partial C_2(a,L)/\partial a}
    \Big)^{1/2} .
\end{equation}
Finally, from (\ref{3HZxxR}) we obtain the apparent toughness of the metamaterial at the mesoscale as
\begin{equation} \label{p6jZjH}
    K_c(a,L)
    =
    \sqrt{E_0' \, g^2(a,L) \bar{\rho} G_c}
    =
    g(a,L) K_{c,0}
\end{equation}
which identifies $g(a,L)$ as a {\sl shielding factor} analogous to those that arise in the analysis of crack-tip shielding in composites and other toughening mechanisms (e.~g., \cite{Hutchinson:1987}). In the present case, the shielding factor $g(a,L)$ is a structural property that accounts for the effect of cell size on the apparent toughness of the metamaterial.

In order to make contact with the experimental data reported in \cite{Shaikeea2022}, we assume bars of circular cross-section, whereupon
\begin{equation}
    A = \pi R^2,
    \quad
    I = \frac{\pi R^4}{4} ,
    \quad
    J = \frac{\pi R^4}{2} ,
\end{equation}
in terms of the radius $R$ of the cross-section, and we change parametrization to
\begin{equation}
    A = \frac{\bar{\rho} L^2}{12\sqrt{2}} ,
    \quad
    I
    =
    \frac{A^2 }{4\pi}
    =
    \frac{\bar{\rho}^2 L^4}{1152\pi} ,
    \quad
    \lambda^2 = \frac{48\sqrt{2}\pi}{\bar{\rho}} ,
\end{equation}
where we have used (\ref{UCgR4X}) and (\ref{Y9zSnV}).
In this parametrization, for small $\bar{\rho}$ the shielding factor (\ref{q2BzAw}) takes the form
\begin{equation}
    g(a\to+\infty,L)
    =
    \sqrt{1 - \frac{2 L^2}{3 h^2}}
    +
    \frac{127 L^2 }{96 \sqrt{6} \pi  h \sqrt{3 h^2-2 L^2}}
    \bar{\rho}
    +
    O(\bar{\rho}^2) ,
\end{equation}
and the apparent toughness (\ref{p6jZjH}) becomes
\begin{equation}
    \frac{K_c(a\to+\infty,L)}{\sqrt{E G_c}}
    \sim
    \frac{\sqrt{3 h^2-2 L^2}}{2 \sqrt{6} h}
    \bar{\rho}
    +
    \frac{ \left(14 h^2+33 L^2\right)}{128 \sqrt{3}
    \pi  h \sqrt{3 h^2-2 L^2}}
    \bar{\rho}^2
    +
    O(\bar{\rho}^3) ,
\end{equation}
where we have used (\ref{Vjm69U}) and (\ref{s2SThn}), and we have taken the limit of $a \to +\infty$ and thus excluded short-crack effects for simplicity.

\begin{figure}[ht!]
\begin{center}
    \includegraphics[width=0.65\textwidth]{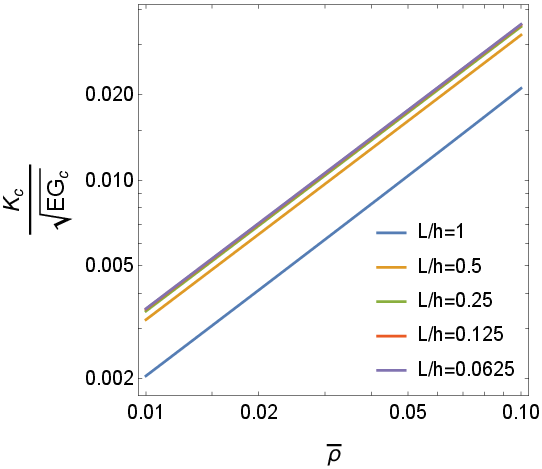}
    \caption{Second-order homogenized octet frame. Dependence of normalized apparent mode-I toughness $K_c(a,L)/\sqrt{E G_c}$ on relative density $\bar{\rho}$ and size $L$ of the unit cell for long cracks $a\to+\infty$.} \label{chEa86}
\end{center}
\end{figure}

The dependence of the normalized apparent mode-I toughness $K_c(a,L)$ for long cracks $a\to+\infty$ on the relative density $\bar{\rho}$ and the size $L$ of the unit cell is shown in Fig.~\ref{chEa86}. As expected, the fully-homogenized value $K_c(a,0)=K_{c,0}$, is recovered in the limit $L\to 0$ of vanishing cell size. At the mesoscale, with no strict separation of scales, the effect of the discreteness of the metamaterial, as measured by the cell size $L$ relative to the size $h$ of the specimen, is to {\sl reduce} the apparent toughness of the metamaterial, i.~e., {\sl coarser is weaker}. However, it should be noted that this reduction in apparent toughness is the result of a structural {\sl crack-tip anti-shielding} effect, i.~e., an increase in the driving force for crack-growth initiation in finite cell size metamaterials relative to that of the fully homogenized metamaterial. Remarkably, the predicted trend of a decrease in apparent toughness with increasing coarsening of the metamaterial is consistent with the experimental data reported in \cite{Shaikeea2022}.

\section{Summary and conclusions}

We have resorted to discrete-to-continuum methods from calculus of variations \cite{Alicandro:2004, Braides:2004} to evaluate the asymptotic behavior of fine metamaterials as a function of cell size. The analysis shows that, to zeroth order in the cell size, the continuum limit of metamaterial frames, undergoing both axial and bending deformations, is {\sl micropolar} \cite{Ariza:2024}, in the sense of \cite{Eringen:1968}, but exhibits linear-elastic fracture mechanics scaling and therefore no size effect. In order to retain information about the unit-cell size when passing to the continuum limit, we extend the asymptotic analysis to higher order by the method of $\Gamma$-expansion \cite{Anzellotti:1993}.

As an application, we evaluate the compliance of double-cantilever octet specimens to second order and resort to compliance analysis to elucidate the dependence of the apparent toughness of the specimen on cell size. The analysis predicts the discreteness of the metamaterial lattice to effectively shield the crack-tip, a mechanism that we term {\sl lattice shielding}. Furthermore, the theory specifically predicts {\sl anti-shielding}, i.~e., {\sl coarser is weaker}, in agreement with the experimental observations of \cite{Shaikeea2022}.

In closing, we note that the ability of the analysis to supply closed-form solutions for quantities of interest, such as the energy-release rate and apparent toughness, as a function of key parameters, such as domain and cell sizes, is quite remarkable. Since many of the observations pertaining to metamaterials at the mesoscale, where strict separation of scales fails, are likely to have a strong structural component, with conditions for failure controlled by the attendant configurational driving forces, the ability to express such structural relations analytically and in closed form greatly enhances intuition and understanding of the phenomena. In this sense, analytical methods such as presented here suggest themselves as worthwhile complements to experimental and computational approaches.

These attractive features notwithstanding, a number of extensions and additional applications of the present method of analysis could also be achieved by recourse to numerical simulation, which could enable consideration of complex 2D lattices~\cite{Fleck:2007}, 3D lattices~\cite{Shaikeea2022}, and others~\cite{shaikeea2024anomalous}, not directly amenable to analytical treatment. The role of bending effects in determining fracture toughness~\cite{berkache2022micropolar} could also be studied systematically for bending-dominated structures, such as hexagonal lattices.

\section*{Acknowledgements}

M.~Ortiz gratefully acknowledges the support of the Deutsche Forschungsgemeinschaft (DFG, German Research Foundation) {\sl via} project 211504053 - SFB 1060; project 441211072 - SPP 2256; and project 390685813 -  GZ 2047/1 - HCM. M.~P.~Ariza gratefully acknowledges financial support from Ministerio de Ciencia e Innovación under grant number PID2021-124869NB-I00. J.~Ulloa and J.~E.~Andrade acknowledge the support from the US ARO MURI program with Grant No. W911NF-19-1-0245.

\begin{appendix}

\section{The discrete Fourier transform} \label{TwWqmj}

Let $(a_i)_{i=1}^n$ be a basis of $\mathbb{R}^n$ and $\mathcal{L} = \{ x(l) = \sum_{i=1}^n l^i a_i  :  l \in \mathbb{Z}^n \}$ the corresponding Bravais lattice. Let $f : \mathcal{L} \to \mathbb{R}$ be a real-valued lattice function. The discrete Fourier transform of $f$ is a complex function $\hat{f}(k)$ supported on the Brillouin zone $B$ in dual space given by
\begin{equation} \label{IYOTsT}
    \hat{f}(k)
    =
    V \sum_{l \in \mathbb{Z}^n}
    f(l) {\rm e}^{-i k \cdot x(l) } ,
\end{equation}
where $V$ is the volume of the unit cell of the lattice. The inverse mapping is given by
\begin{equation} \label{aX0TEN}
    f(l)
    =
    \frac{1}{(2\pi)^n} \int_B
    \hat{f}(k) {\rm e}^{i k \cdot x(l) } dk .
\end{equation}
The convolution of two lattice functions $f(l)$, $g(l)$ is
\begin{equation} \label{YNU3CC}
    (f*g)(l) = V \sum_{l' \in \mathbb{Z}^n} f(l-l') g(l') ,
\end{equation}
whereupon the convolution theorem states that
\begin{equation} \label{m4IeYG}
    \widehat{f*g} = \hat{f} \hat{g} .
\end{equation}
In addition, the Parseval identity states that
\begin{equation} \label{pB0rWJ}
    V \sum_{l \in \mathbb{Z}^n} f(l) g^*(l)
    =
    \frac{1}{(2\pi)^n} \int_{B}
    \hat{f}(k) \hat{g}^*(k) dk
\end{equation}
which establishes an isometric isomorphism between $l^2$ and $L^2(B)$ (cf.~\cite{Ariza:2005} and references therein for further details).
\end{appendix}

\end{document}